\documentclass[12pt,a4paper,oneside]{article}
\usepackage{graphicx,amssymb,amsmath,color}
\usepackage[linkcolor={blue},citecolor={red},colorlinks=true]{hyperref}
\usepackage[all]{hypcap} 
\usepackage{cite}
\usepackage{relsize}
\usepackage{enumerate}
\usepackage[margin=3 cm]{geometry}
\usepackage{slashed}
\usepackage[normalem]{ulem}
\usepackage[dvipsnames, x11names]{xcolor}
\usepackage{xcolor}
\usepackage{mathtools}
\usepackage{lipsum}

\newcommand{\be}{\begin{equation}}
\newcommand{\ee}{\end{equation}}

\def\beq{\begin{equation}}
\def\eeq{\end{equation}}
\def\bea{\begin{eqnarray}}
\def\eea{\end{eqnarray}}

\def\bit{\begin{itemize}}
\def\eit{\end{itemize}}

\def\baa{\begin{array}}
\def\eaa{\end{array}}

\newcommand{\dd}{\mathrm{d}}

\newcommand{\mah}{m_{a,\mathrm{h}}}
\newcommand{\mbh}{m_{b,\mathrm{h}}}
\newcommand{\mch}{m_{c,\mathrm{h}}}
\newcommand{\mas}{m_{a,\mathrm{s}}}
\newcommand{\mbs}{m_{b,\mathrm{s}}}
\newcommand{\mcs}{m_{c,\mathrm{s}}}
\newcommand{\mai}{m_{a,\mathrm{in}}}
\newcommand{\mao}{m_{a,\mathrm{out}}}

\newcommand{\pref}[1]{(\ref{#1})}
\newcommand{\eref}[1]{eq.~(\ref{#1})}

\newcommand{\fref}[1]{fig.~\ref{#1}}
\newcommand{\Fref}[1]{Fig.~\ref{#1}}

\newcommand{\Rref}[1]{Ref.~\cite{#1}}

\newcommand{\Rrefs}[1]{Refs.~\cite{#1}}

\definecolor{chromeyellow}{rgb}{1.0, 0.65, 0.0}
\definecolor{darkcoral}{rgb}{0.8, 0.36, 0.27}
\definecolor{cadmiumgreen}{rgb}{0.0, 0.42, 0.24}

\begin{document}

\begin{flushright}
DESY-25-163
\end{flushright}
\vspace{.6cm}
\begin{center}

{\Large \bf 
Bubble Friction in Symmetry-Restoring Transitions
\\}

\vspace{1cm}{Andrew J. Long,$^{1}$ \ Bibhushan Shakya,$^{2,3}$ \ and \ Julia Anabell Ziegler\,$^{4}$}
\\[5mm]

{$^1$ \it Department of Physics and Astronomy, Rice University, Houston, Texas 77005, U.S.A.}\\

{$^2$ \it William H.\ Miller III Department of Physics \& Astronomy, Johns Hopkins University, 3400 N.\ Charles St., Baltimore, MD 21218, USA}\\

{$^3$ \it Deutsches Elektronen-Synchrotron DESY, Notkestr.\,85, 22607 Hamburg, Germany}\\

{$^4$ \it II. Institute of Theoretical Physics, Universität Hamburg, Luruper Chaussee 149, 22761, Hamburg, Germany}
\end{center}

\bigskip \bigskip \bigskip

\centerline{\bf Abstract} 
\begin{quote}
In standard (symmetry-breaking) first-order phase transitions, the frictional pressure on expanding bubble walls can be dominated by transition radiation -- the emission of a gauge boson with phase-dependent masses as particles present in the thermal plasma pass through bubble walls. This process is enhanced in the soft limit, and is known to produce a significant frictional effect that is proportional to the Lorentz factor $\gamma$ of the bubble wall, thereby prohibiting runaway behavior.  We calculate the analogous pressure for phase transitions with symmetry restoration. In such transitions, we show that the pressure due to this process can be $\textit{negative}$, producing the opposite effect. However, when the Lorentz factor of the wall gets very large, the result approaches the same scaling as the standard scenarios. Therefore, phase transitions with symmetry restoration can feature an intermediate negative friction regime even in the presence of significant interactions with the plasma, and the bubble wall terminal Lorentz factor can be significantly larger (by more than an order of magnitude) than in the corresponding symmetry-breaking scenarios. This can carry important implications for various phenomenological applications, from gravitational waves to physics beyond-the-Standard-Model.
\end{quote}

\newpage

\tableofcontents

\section{Introduction}
\label{sec:intro}

First-order phase transitions (FOPTs) have received significant research scrutiny in recent years, driven primarily by the prospect that they are well-motivated early Universe events in several beyond-the-Standard-Model (BSM) frameworks that can produce observable stochastic gravitational wave (GW) signals. 

An important parameter in the phenomenology of FOPTs is the speed of the bubble walls of the stable vacuum. This parameter is known to be of particular importance in models of electroweak baryogenesis (see e.g.~\cite{Barrow:2022gsu} for some recent overviews), where the interactions between the expanding bubble walls and the surrounding plasma is responsible for creating a nonvanishing baryon asymmetry, with the amount of asymmetry produced being sensitive to the wall speed. Understanding whether the bubble walls achieve some terminal velocity or continue to accelerate (the so-called runaway regime) is also important for deducing the form of the GWs produced from such transitions. If the walls run away, most of the energy released in the phase transition is stored in the bubble walls, and GWs are sourced by the scalar field energy densities during bubble collision~\cite{Kosowsky:1991ua,Kosowsky:1992rz,Kosowsky:1992vn,Kamionkowski:1993fg,Caprini:2007xq,Huber:2008hg,Bodeker:2009qy,Jinno:2016vai,Jinno:2017fby,Konstandin:2017sat,Cutting:2018tjt,Cutting:2020nla,Lewicki:2020azd} or by the distribution of particles produced from bubble collisions~\cite{Inomata:2024rkt}. On the other hand, in the presence of significant energy transfer from the walls to the particles surrounding the bubbles, most of the energy gets transferred to these particles, and GWs instead arise from sound waves~\cite{Hindmarsh:2013xza,Hindmarsh:2015qta,Hindmarsh:2017gnf,Cutting:2019zws,Hindmarsh:2016lnk,Hindmarsh:2019phv} and turbulence~\cite{Kamionkowski:1993fg,Caprini:2009yp,Brandenburg:2017neh,Cutting:2019zws,RoperPol:2019wvy,Dahl:2021wyk,Auclair:2022jod} in the plasma, or through nontrivial spatial configurations of particles that do not thermalize~\cite{Jinno:2022fom}. Moreover, the speed (and consequently the energy density) of the bubble wall also determines the energy scale and efficiency of producing high energy or high mass particles from the bubble walls interacting with the plasma or with other bubbles, which has found many applications for various BSM phenomena such as dark matter production and baryogenesis~\cite{Azatov:2021ifm,Baldes:2021vyz,Azatov:2021irb,Baldes:2022oev,Chun:2023ezg,Mansour:2023fwj,Shakya:2023kjf,Ai:2024ikj,Giudice:2024tcp,Cataldi:2024pgt,Cataldi:2025nac}. 

The bubble wall speed is determined by the interactions between the bubble walls and the particles present in the surrounding plasma, and was studied in detail in the seminal papers by B\"odeker and Moore~\cite{Bodeker:2009qy,Bodeker:2017cim} for symmetry-breaking phase transitions. At leading order, the plasma imparts some pressure on the expanding bubble walls as the particles crossing into the bubbles become massive in the broken phase inside the bubbles. This pressure was calculated to be independent of the speed of the bubble walls. At next-to-leading order (NLO), a different process becomes crucial: transition radiation, the emission of a vector boson as particles cross into bubble walls, was shown to produce a pressure that scales linearly with the Lorentz factor of the bubble wall, which inevitably saturates the latent energy released from the vacuum and leads to a terminal velocity for the bubble wall~\cite{Bodeker:2017cim}. The dominant effect comes from the emission of soft bosons, which has an enhanced probability and imparts the largest pressure on the bubble walls. While these calculations were performed with several approximations, other subsequent, more rigorous calculations in the literature have confirmed the general veracity of these results~\cite{Hoche:2020ysm,Gouttenoire:2021kjv,Ai:2024shx,Long:2024sqg,Ai:2025bjw}.

While the above results hold for symmetry-breaking transitions, which are the primary form of FOPTs studied in the literature, there has recently been growing interest in FOPTs with symmetry \textit{restoration}. Such transitions could generally take place, in particular, during reheating or preheating~\cite{Buen-Abad:2023hex}, including for the Standard Model Higgs \cite{Shakya:2025mdh}  (see \Rrefs{Barni:2025gnm,Barni:2025mud,Suematsu:2017iki} for other studies). In this paper, we study the nature of plasma friction on the expanding bubble walls for such symmetry-restoring transitions, as such configurations display several qualitatively different characteristics compared to the standard symmetry-breaking transitions. First, the leading order pressure from particles crossing into the bubbles in this case is $\textit{negative}$, since particles are massive outside the bubbles in the broken phase but lose mass when entering the symmetry restored phase inside the bubbles, thereby creating an $\textit{antifriction}$ effect. As we demonstrate in this paper, the NLO pressure arising from the emission of gauge bosons can also be negative, especially in the soft limit that dominates the distribution, producing the same antifrictional effect. 

On the other hand, a recent calculation~\cite{Azatov:2024auq} demonstrated that the NLO pressure in symmetry-restoring phase transitions scales in the same manner as in the symmetry-breaking scenario (i.e, positive pressure that grows linearly with the wall Lorentz factor) in the limit of large Lorentz factor. Given the broad phenomenological importance of the bubble wall speed, it then becomes important to understand how the pressure transitions from a negative effect to a positive one, restoring the standard behavior known from symmetry-breaking transitions, and how it impacts the terminal velocity achieved by the bubble walls. In this paper, we attempt to answer these questions by extending the original calculations of B\"odeker and Moore~\cite{Bodeker:2009qy,Bodeker:2017cim}, which were performed for the symmetry-breaking scenario, to the symmetry-restoring scenario.\footnote{Note that the calculation in \Rref{Azatov:2024auq} cannot answer these questions. This paper demonstrated that the standard (symmetry-breaking) result re-emerges in the large Lorentz factor limit, but did not calculate at what value it is reached. In fact, \Rref{Azatov:2024auq} speculated that an intermediate regime of negative friction might be possible.} This will inherit the simplifying approximations used in these calculations, but should provide a correct order-of-magnitude estimate of the effect, and shed light on the underlying physics.

\section{Thermal pressure}
\label{sec:thermal_pressure}

\begin{figure}[t]
    \centering
    \includegraphics[width=0.48\linewidth]{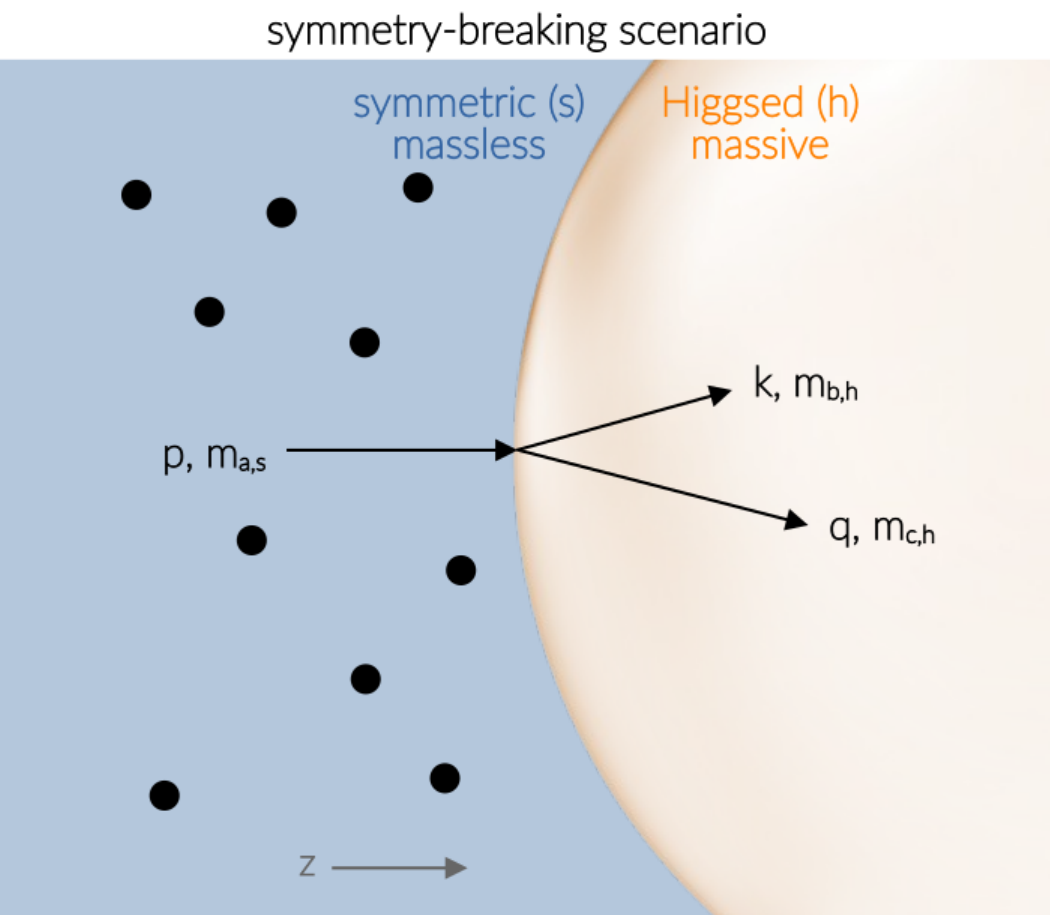} \hfill 
    \includegraphics[width=0.48\linewidth]{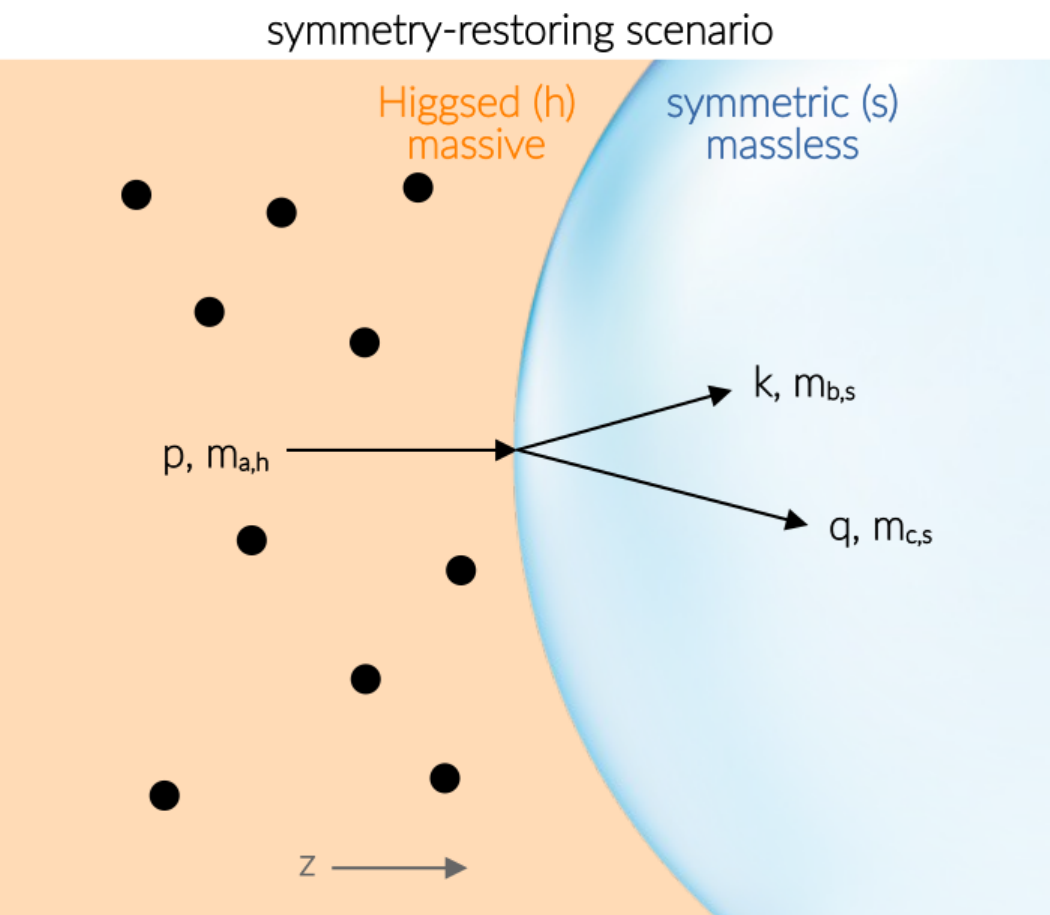}
    \caption{\label{fig:sketches}
    Illustration of the 1-to-2 transition, the main process studied in this paper.  \textit{Left:}  For the symmetry-breaking scenario, massless particles ($\mas = 0$) are incident on the bubble wall from the symmetric phase, and they transition into pairs of massive particles ($\mbh, \mch > 0$) in the broken phase.  \textit{Right:}  For the symmetry-restoring scenario, the phases are switched, so that incident particles are massive and emergent particles are massless.  In the rest frame of the plasma, the wall moves left with Lorentz factor $\gamma$; in the rest frame of the wall, the particles have a thermal distribution of momenta boosted to the right by Lorentz factor $\gamma$.  
    }
\end{figure}

Thermal pressure arises when particles in the plasma move across the bubble wall and experience a change in mass.  
We contrast the symmetry-breaking scenario (increasing mass) against the symmetry-restoring scenario (decreasing mass). 
We denote the two phases as symmetric phase (subscript $s$) and broken phase (subscript $h$ for Higgsed).  
Particles in the symmetric phase are assumed to be massless, i.e. $\mas = \mbs = \mcs = 0$, and particles in the broken phase have masses denoted by $\mah$, $\mbh$, $\mch$, and so on.\footnote{We have verified that the results derived in this paper remain applicable if the particles have nonzero but small masses in the symmetric phase, on the order of $m_\mathrm{s} = 0.1 m_\mathrm{h}$.} 
For the symmetry-breaking scenario, the broken phase is inside the bubble, and for the symmetry-restoring scenario, the broken phase is outside the bubble.  
These two scenarios are illustrated in Fig.~\ref{fig:sketches}.

Throughout this work we restrict our attention to bubble walls that are moving at relativistic velocities relative to the plasma.  
If $\mathbf{v}_w$ denotes the wall's velocity in the rest frame of the plasma and $\gamma = 1 / \sqrt{1 - |\mathbf{v}_w|^2}$ is the associated Lorentz factor, then our regime of interest corresponds to $\gamma \gg 1$.  
The calculation of the thermal pressure is significantly simplified for relativistic walls~\cite{Bodeker:2009qy}.  
Since the wall travels much faster than the speed of sound ($c_s \approx 1 / \sqrt{3}$ for a relativistic plasma), nearly every particle passes into the bubble with negligible reflections.  
Consequently, the plasma in front of the wall is unaffected by the wall's approach, and the flux of particles onto the wall is simply a thermal distribution.  

The thermal pressure $\mathcal{P}_\mathrm{th}$ may be calculated using perturbation theory as a series in powers of squared couplings, $\mathcal{P}_\mathrm{th} = \mathcal{P}_{1\to1} + \mathcal{P}_{1\to2} + \cdots$.  
The leading term, called the one-to-one pressure and denoted by $\mathcal{P}_{1\to1}$, is zeroth order in the squared couplings~\cite{Bodeker:2009qy}.  
The next-to-leading term, called the one-to-two pressure and denoted by $\mathcal{P}_{1\to2}$, is first-order in the squared couplings, and it captures the effect of transition radiation~\cite{Bodeker:2017cim}.  
See \Rrefs{Espinosa:2010hh,Dorsch:2018pat,Hoche:2020ysm,Gouttenoire:2021kjv,Ai:2024shx,Long:2024sqg,Ai:2025bjw} for studies of thermal pressure on relativistic bubbles.  
In the remainder of this section, we summarize results from this literature for 1-to-1 and 1-to-2 contributions to the thermal pressure, emphasizing differences between the symmetry-breaking and restoring scenarios.  

\subsection{1-to-1 pressure}

Consider a particle that has mass $\mas = 0$ in the symmetric phase and mass $\mah$ in the broken phase.  
For the symmetry-breaking scenario, the mass rises from $m_{a,\mathrm{out}} = \mas = 0$ to $m_{a,\mathrm{in}} = \mah > 0$ when the particle enters the bubble.  
For the symmetry-restoring scenario, the mass decreases from $m_{a,\mathrm{out}} = \mah > 0$ to $m_{a,\mathrm{in}} = \mas = 0$.  
We work in the rest frame of the wall, and we treat the wall as planar since its curvature radius is significantly larger than particle physics length scales. 
The background of the planar wall at rest maintains time-translation invariance and transverse spatial-translation invariance (along the wall), but it is not invariant under longitudinal spatial translations (normal to the wall).  
As a result, a particle may cross the wall while maintaining its energy $E$ and transverse momentum $\mathbf{p}_\perp$, but while losing or gaining longitudinal momentum $p_z$ to compensate for the increase or decrease of its mass.  


The change in the particle's longitudinal momentum ($z$-component) is calculated as~\cite{Bodeker:2009qy}
\begin{equation}
\begin{split}
    & E_a^2 = p_{a,z,\mathrm{in}}^2 + \mathbf{p}_{a,\perp}^2 + \mai^2 = p_{a,z,\mathrm{out}}^2 + \mathbf{p}_{a,\perp}^2 + \mao^2 \\ 
    & \ \Rightarrow \ p_{a,z,\mathrm{in}} - p_{a,z,\mathrm{out}} = \frac{\mao^2 - \mai^2}{p_{a,z,\mathrm{in}} + p_{a,z,\mathrm{out}}} \approx \frac{\mao^2 - \mai^2}{2 E_a}
    \;.
\end{split}
\end{equation}
The approximation follows because the particle is boosted in the longitudinal direction (in the rest frame of the wall). 
The wall's longitudinal momentum change, denoted by $\Delta p_{1\to1}$, is opposite to the particle's longitudinal momentum change.  
Therefore, we can evaluate 
\begin{align}\label{eq:Dp_1to1}
    \Delta p_{1\to1} \equiv p_{a,z,\mathrm{out}} - p_{a,z,\mathrm{in}} \approx \frac{m_{a,\mathrm{in}}^2 - m_{a,\mathrm{out}}^2}{2E_a} 
    = \begin{cases} 
    + \tfrac{1}{2E_a} \mah^2 & , \quad \text{sym-breaking scenario} \\ 
    - \tfrac{1}{2E_a} \mah^2 & , \quad \text{sym-restoring scenario} 
    \end{cases} 
    \;.
\end{align}
For the symmetry-breaking scenario, $m_{a,\mathrm{in}} = \mah$ and $m_{a,\mathrm{out}} = \mas = 0$; hence, the wall's longitudinal momentum change $\Delta p_{1\to1} \approx \mah^2 / 2E_a$ is positive.  
For the symmetry-restoring scenario, $m_\mathrm{out} = \mah$ and $m_\mathrm{in} = \mas = 0$; hence, the wall's longitudinal momentum change $\Delta p_{1\to1} \approx -\mah^2 / 2E_a$ is negative.  
In the rest frame of the plasma, the wall moves in the negative $z$-direction, so the symmetry-breaking scenario's positive $z$-momentum change implies a deceleration, and the symmetry-restoring scenario's negative $z$-momentum change implies an acceleration.

This opposite behavior carries over to the thermal pressure. The thermal pressure is calculated by integrating the wall's longitudinal momentum change over the thermal flux of incident particles~\cite{Arnold:1993wc}.
Let $f_a(\mathbf{p})$ denote the thermal one-particle phase space distribution function of the incident $a$-type particles in the rest frame of the wall.  
Then the pressure is calculated by integrating~\cite{Bodeker:2009qy} 
\be\label{eq:P_1to1}
    \mathcal{P}_{1 \to 1} 
    = \int \! \frac{\dd^3 p_{a,\mathrm{s}}}{(2\pi)^3} \, \frac{p_{a,z,\mathrm{s}}}{E_a} \, f_a(\mathbf{p}_{a,\mathrm{s}}) \, \Delta p_{1 \to 1}
    \approx \begin{cases} 
    + \frac{1}{24} \mah^2 T^2 & , \quad \text{sym-breaking scenario} \\ 
    - \frac{1}{24} \mah^2 T^2 & , \quad \text{sym-restoring scenario} 
    \end{cases} 
    \;.
\ee
For the symmetry-breaking scenario the wall experiences a positive thermal pressure, which is friction force that retards its motion~\cite{Bodeker:2009qy}.  
For the symmetry-restoring scenario the wall experiences a negative thermal pressure, which is an ``antifriction'' force that accelerates its motion~\cite{Buen-Abad:2023hex}.  

If the full thermal pressure were well-approximated by the 1-to-1 contribution, $\mathcal{P}_\mathrm{th} \approx \mathcal{P}_{1\to1}$, then bubble walls in the symmetry-restoring scenario would achieve runaway behavior~\cite{Buen-Abad:2023hex}.  
That is to say, a wall's speed would approach $v_w = 1$ and its Lorentz factor would grow arbitrarily large $\gamma \to \infty$.  
Next we will discuss how the 1-to-2 contribution $\mathcal{P}_{1\to2}$ dominates over the 1-to-1 contribution at large wall speeds, and prohibits a runaway~\cite{Azatov:2024auq}. 

\subsection{1-to-2 pressure}

Interactions allow for different behaviors when particles encounter the bubble wall~\cite{Bodeker:2017cim}.  
Consider three species of particles having masses $\mas = \mbs = \mcs = 0$ in the symmetric phase and $\mah$, $\mbh$, $\mch \geq 0$ in the broken phase.  
When a particle of species $a$ is incident on the wall, there is a nonzero probability for it to convert into two particles, one each of species $b$ and $c$, that pass into the bubble.  
The wall's longitudinal momentum change is calculated as 
\begin{align}\label{eq:Dp_1to2}
    \Delta p_{1\to2} 
    \equiv p_{z,\mathrm{out}} - k_{z,\mathrm{in}} - q_{z,\mathrm{in}} 
    \approx \frac{1}{2E_a} \biggl( - m_{a,\mathrm{out}}^2 + \frac{m_{b,\mathrm{in}}^2}{x} + \frac{m_{c,\mathrm{in}}^2}{1-x} + \frac{k_\perp^2}{x(1-x)} \biggr) 
    \;,
\end{align}
where $\mathbf{p}_\mathrm{out}$ is the momentum of the incident $a$-particle, and where $\mathbf{k}_\mathrm{in}$ and $\mathbf{q}_\mathrm{in}$ are the momenta of the recoiling $b$- and $c$-particles, respectively. 
The approximation uses $E_a = E_b + E_c$, $x \equiv E_b / E_a$, $\mathbf{p}_\perp = \mathbf{k}_\perp + \mathbf{q}_\perp$, $k_\perp \equiv |\mathbf{k}_\perp|$, $\mathbf{p}_\perp \approx \mathbf{0}$, $p_{z,\mathrm{out}} \gg m_{a,\mathrm{out}}, k_\perp$, $k_{z,\mathrm{in}} \gg m_{b,\mathrm{in}}, k_\perp$, and $q_{z,\mathrm{in}} \gg m_{c,\mathrm{in}}, k_\perp$. 
In what follows, we will additionally assume that the $b$-particle is soft, so that its energy $E_b = x E_a \ll E_a$ is much smaller than the incident $a$-particle's energy. 

For the symmetry-breaking scenario $\mathrm{out} = s$ and $\mathrm{in} = h$, and for the symmetry-restoring scenario $\mathrm{in} = s$ and $\mathrm{out} = h$.  
The wall's longitudinal momentum change is given by 
\begin{align}\label{eq:Dp_1to2_cases}
    \Delta p_{1\to2} \approx \frac{1}{2E_a} \begin{cases} 
    \frac{\mbh^2}{x} + \frac{\mch^2}{1-x} + \frac{k_\perp^2}{x(1-x)} & , \quad \text{sym-breaking scenario} \\ 
    - \mah^2 + \frac{k_\perp^2}{x(1-x)} & , \quad \text{sym-restoring scenario} \end{cases} 
    \;.
\end{align}
For the symmetry-breaking scenario, the wall's longitudinal momentum change $\Delta p_{1\to2}$ is positive, which is the same sign as $\Delta p_{1\to1}$.  
However, for the symmetry-restoring scenario, the wall's longitudinal momentum change $\Delta p_{1\to2}$ may be either positive or negative depending on the kinematic variables $k_\perp$ and $x$.  
This is crucially different from the 1-to-1 calculation \pref{eq:Dp_1to1} where $\Delta p_{1\to1}$ is always negative for the symmetry-restoring scenario.  
Note that $\Delta p_{1\to2} < 0$ for $k_\perp < \mah [x(1-x)]^{1/2}$.  
It follows that the thermal pressure $\mathcal{P}_{1\to2}$ is negative for small values of the Lorentz factor $\gamma$, and it is positive for large $\gamma$.  
Roughly speaking, this is because the phase space integral over $k_\perp$ is dominated by the region $k_\perp\sim \mbh$ and the $x-$ integral is dominated by the IR cutoff $x_\mathrm{min} = \mbh/E_a$, and we take $E_a = \gamma T$ such that $k_\perp^2 / x \sim \mbh \gamma T$.  
Thus for large $\gamma$ the phase space integrals are dominated by the positive terms, giving $\Delta p_{1\to2} > 0$ for both the symmetry-breaking and restoring scenarios.  
This is consistent with the results of \Rref{Azatov:2024auq}, whose authors found that the pressure in the symmetry-restoring scenario approaches its counterpart in the symmetry-breaking scenario in the limit of large Lorentz factor. 

The thermal pressure is calculated by the integral~\cite{Bodeker:2017cim} (see also Sec.~3B of \Rref{Hoche:2020ysm}) 
\begin{align}\label{eq:P_1to2}
    \mathcal{P}_{1 \to 2}
    &= \int \! \frac{\dd^3 p_\mathrm{s}}{(2\pi)^3 2E_a} \int \! \frac{\dd^3 k_\mathrm{s}}{(2\pi)^3 2E_b} \int \! \frac{\dd^3 q_\mathrm{s}}{(2\pi)^3 2E_c} \, \frac{p_{z,\mathrm{s}}}{E_a} f_a(\mathbf{p}_\mathrm{s}) \bigl( 1 \pm f_b(\mathbf{k}_\mathrm{s}) \bigr) \bigl( 1 \pm f_c(\mathbf{q}_\mathrm{s}) \bigr) 
    \\
    & \quad \times (2\pi)^3 \delta^2 (\textbf{p}_\perp - \textbf{k}_\perp - \textbf{q}_\perp) \, \delta(E_a - E_b - E_c) \, |\mathcal{M}|^2 \, \Delta p_{1\to2}  
    \;. 
    \nonumber
\end{align}
For small occupation numbers we approximate $(1 \pm f_{b,c}) \approx 1$. 
The matrix element $\mathcal{M}$ was calculated by the authors of \Rref{Bodeker:2017cim} using the WKB approximation as
\beq
   |\mathcal{M}|^2 
    = 4 E_a^2 \left| \frac{V_\mathrm{h}}{A_\mathrm{h}} - \frac{V_\mathrm{s}}{A_\mathrm{s}} \right|^2
    \;,
    \label{eq:matrix}
\eeq
where $V_\mathrm{s}$ and $V_\mathrm{h}$ are the vertex functions in the symmetric and broken phases, respectively. 
We focus on the emission of a transversely polarized vector boson\footnote{The transverse polarization modes are conceptually and technically easier to study than the longitudinal polarization mode.  Whereas the vector boson's transverse polarizations are present in both the symmetric and broken phases, its longitudinal polarization is only present in the broken phase.  The authors of \Rref{Azatov:2023xem} found that both polarizations contribute comparably to the pressure for symmetry-breaking transitions with mild super-cooling. Therefore, we don't expect a significant error by neglecting the longitudinal polarization.
} 
with energy $E_b$ and transverse momentum $\mathbf{k}_\perp$ for which (see Tab.~1 in \Rref{Bodeker:2017cim}) 
\beq\label{eq:V_sq_transverse}
    |V_\mathrm{s}|^2 
    = |V_\mathrm{h}|^2 
    \approx 4 g^2 C_2[R] \frac{1}{x^2} k_\perp^2
    \;, 
\eeq
where $g$ is the gauge coupling constant, $C_2[R]$ is the second Casimir, $x = E_b / E_a$ is assumed to be $0 \leq x \ll 1$, and $k_\perp \equiv |\mathbf{k}_\perp|$.  
The kinematical factors are given by\footnote{We have verified that the impact of these approximations on the thermal pressure is sub-$\mathcal{O}(10\%)$. } 
\begin{equation}
\begin{split}
    A_\mathrm{s} 
    & = - 2 E_a \bigl( p_{z,\mathrm{s}} - k_{z,\mathrm{s}} - q_{z,\mathrm{s}} \bigr) 
    \approx \mas^2 - \frac{\mbs^2}{x} - \frac{\mcs^2}{1-x} - \frac{k_\perp^2}{x(1-x)}\,, \\ 
    A_\mathrm{h} & = - 2 E_a \bigl( p_{z,\mathrm{h}} - k_{z,\mathrm{h}} - q_{z,\mathrm{h}} \bigr) 
    \approx \mah^2 - \frac{\mbh^2}{x} - \frac{\mch^2}{1-x} - \frac{k_\perp^2}{x(1-x)} 
    \;.
\end{split}
\end{equation}
where we make the same approximations as in \eref{eq:Dp_1to2}. 
Putting together these factors gives the squared matrix element~\cite{Bodeker:2017cim} (see also App.~B in \Rref{Long:2024sqg}) 
\beq\label{eq:Msq}
   |\mathcal{M}|^2 
    = 16 g^2 C_2[R] \frac{k_\perp^2 \, \bigl( \mbh^2 - \mbs^2 \bigr)^2}{\bigl( k_\perp^2 + \mbh^2 \bigr)^2 \bigl( k_\perp^2 + \mbs^2 \bigr)^2} E_a^2 
    \;,
\eeq
where we have kept only the terms that are leading as $x \to 0$.  
Note that this expression is identical for both the symmetry-breaking scenario (with $m_{b,\mathrm{out}} = \mbs = 0$ and $m_{b,\mathrm{in}} = \mbh = m_b$) and for the symmetry-restoring scenario (with $m_{b,\mathrm{out}} = \mbh = m_b$ and $m_{b,\mathrm{in}} = \mbs = 0$). 
This is to be expected, since the symmetry-breaking and restoring scenarios simply switch the ``in" and ``out" labels, and the squared matrix element in \eref{eq:matrix} is invariant.

Combining the expressions above allows the thermal pressure in \eref{eq:P_1to2} 
to be estimated as\footnote{We only integrate over $k_{z,\mathrm{s}} > 0$ corresponding to particles that pass into the bubble.  Particles with $k_{z,\mathrm{s}} < 0$ would contribute a larger momentum transfer, but the probability is low enough to neglect them.} 
\begin{align}\label{eq:P_1to2_simple}
    \mathcal{P}_{1 \to 2}
    & \approx \frac{g^2 C_2[R] \, T^2}{4\pi^2} 
    \int_{k_{\perp,\mathrm{min}}}^{k_{\perp,\mathrm{max}}} \! \frac{\dd k_\perp}{k_\perp} 
    \int_{x_\mathrm{min}}^{x_\mathrm{max}} \! \frac{\dd x}{x} 
    \frac{\mbh^4}{\bigl( k_\perp^2 + \mbh^2 \bigr)^2} \, \begin{cases} 
    \frac{\mbh^2}{x} + \mch^2 + \frac{k_\perp^2}{x} & , \quad \text{sym-breaking} \\ 
    - \mah^2 + \frac{k_\perp^2}{x} & , \quad \text{sym-restoring} \end{cases}  
    \;. 
\end{align}
For the limits of integration we have introduced infrared (IR) and ultraviolet (UV) cutoffs.  
This is because we have made several approximations when evaluating the integrand, and these approximations are only valid in a restricted domain of integration that gives the largest contribution to the integral.  
For the UV cutoffs we take $k_{\perp,\mathrm{max}} = [x^2 E_a^2 - \mbh^2]^{1/2}$ and $x_\mathrm{max} = 1 - \mch/E_a$ in the symmetry-breaking scenario, and we take $k_{\perp,\mathrm{max}} = x E_a$ and $x_\mathrm{max} = 1$ in the symmetry-restoring scenario. 
For the IR cutoffs we take $k_{\perp,\mathrm{min}} = 0.1 \mbh$ and $x_\mathrm{min} = \mbh / E_a$.  
Various choices of IR cutoffs were discussed by the authors of \Rref{Gouttenoire:2021kjv}, with the thermal mass of $b$ or a phase-space saturation mass considered as viable options. Both of these evaluate to $\sim 0.1 \mbh$, hence we use this as an IR cutoff for $k_\perp$ in our calculations.

\section{Results and Discussions}
\label{sec:results}

In this section we calculate the thermal pressure using numerical integration methods to evaluate the phase space integrals in \eref{eq:P_1to2_simple}.  
We assess how the pressure depends on the bubble wall's velocity via the Lorentz factor $\gamma$, and we contrast the symmetry-breaking and symmetry-restoring scenarios.  
For the symmetry-restoring scenario, we calculate the value of the Lorentz factor where the pressure switches from negative to positive, and where the pressure approaches the same form as in the symmetry-breaking scenario.  
Finally, we also discuss various phenomenological implications of this modified form of pressure. 

In order to demonstrate a comparison of the symmetry-breaking and symmetry-restoring scenarios, we focus on a model in which $a$ and $c$ are the same particle species. 
The pressure above is then determined by five independent parameters: the broken-phase masses $\mah = \mch$ and $\mbh$, the temperature $T$, the Lorentz factor of the bubble wall $\gamma$, and the gauge factor $g^2 \, C_2[R]$. 
We will evaluate dimensionful parameters in units of $T$. 
The scaling with $g$ is trivial, since it only appears as an overall normalization via \eref{eq:V_sq_transverse}. The three nontrivial independent parameters are thus $\mah/T, \mbh/T, \gamma$.  
Note that the expectation value of the scalar Higgs field in the broken phase $v_\mathrm{h}$ only affects the thermal pressure via the relation $\mbh = g v_\mathrm{h}$.  
It also impacts the bubble wall's terminal velocity via the latent vacuum energy, which we discuss in the next subsection. 

\subsection{Pressure comparison}
\label{sub:pressure_comparison}


\begin{figure}[t!]
    \centering
    \includegraphics[width=0.48\linewidth]{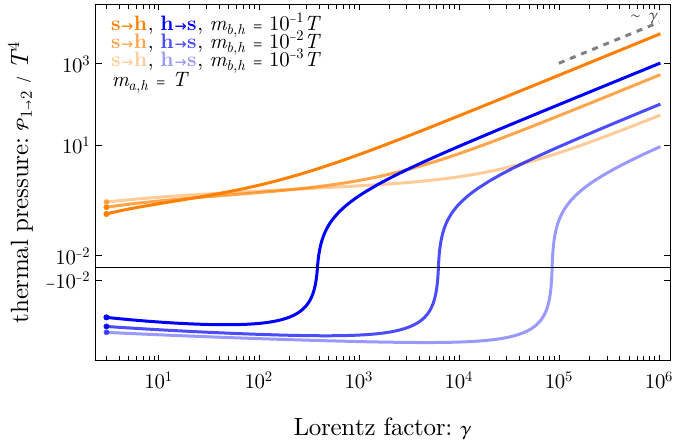} \hfill
    \includegraphics[width=0.48\linewidth]{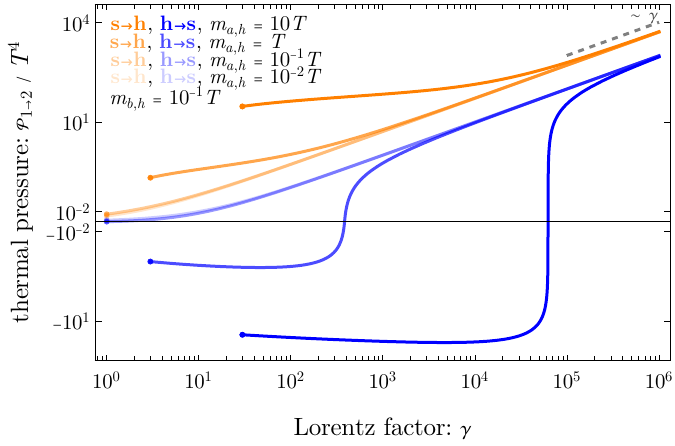}
    \caption{\label{fig:plot_all_contributions}
    Dependence of the thermal pressure on the speed of the bubble wall.  We plot the 1-to-2 thermal pressure $\mathcal{P}_{1 \to 2}$ as a function of the bubble wall's Lorentz factor $\gamma$.  Orange curves labeled $\mathrm{s} \to \mathrm{h}$ indicate the symmetry-breaking scenario, and blue curves labeled $\mathrm{h} \to \mathrm{s}$ indicate the symmetry-restoring scenario.  In the first panel we fix $\mah = \mch = T$ and we vary $\mbh$, and in the second panel we fix $\mbh = 0.1 T$ and we vary $\mah = \mch$.  We take the gauge factors $g^2 C_2[R] = 1$, and more generally the pressure scales linearly with this factor. 
    }
\end{figure}

\Fref{fig:plot_all_contributions} illustrates how the 1-to-2 thermal pressure $\mathcal{P}_{1 \to 2}$ varies with the Lorentz factor $\gamma$ for a few fiducial masses in both the symmetry-breaking and symmetry-restoring scenarios. 
Note that we hold the gauge factor fixed to $g^2 C_2[R] = 1$ while varying $\mbh$. 
If the $b$-particles are gauge bosons that receive their mass from the Higgs field that forms the bubble, then the coupling and mass are related by $\mbh = g v_\mathrm{h}$ where $v_\mathrm{h}$ is the expectation value of the Higgs field inside the bubble, so $v_\mathrm{h} = \mbh / g$ varies implicitly with $\mbh$.  
If the Lorentz factor $\gamma$ becomes too small, then our approximations break down; we thus truncate the curves at $\gamma T < 3 \min[\mah , \mbh]$, indicated by dots in the panels.

By inspecting the figure, we observe several notable features.  
For large values of the Lorentz factor $\gamma$, the 1-to-2 thermal pressure scales as $\mathcal{P}_{1\to2} \propto \gamma$ for both the symmetry-breaking and the symmetry-restoring scenarios and for all the masses shown. 
As noted in earlier studies~\cite{Bodeker:2017cim,Gouttenoire:2021kjv,Ai:2024shx,Ai:2025bjw}, this scaling arises because the average momentum transfer $\langle \Delta p_{1\to2} \rangle \propto \gamma^0$ at large $\gamma$, and the pressure scales with the Lorentz-boosted flux $\propto \gamma^1$.  
Also for large values of $\gamma$, we observe that the pressure scales with the masses as $\mathcal{P}_{1\to2} \propto \mbh^1 \mah^0$.  
One can see from \eref{eq:P_1to2_simple} that the pressure goes as $\mbh^2 / x_\mathrm{min}$ and we take the IR cutoff as $x_\mathrm{min} = \mbh/E_a$.  

Regarding the sign of the 1-to-2 thermal pressure, we note a striking difference between the symmetry-breaking and symmetry-restoring scenarios.  
For the symmetry-breaking scenario, the pressure is positive for all values of $\gamma$, which implies that the thermal pressure always tends to retard the motion of the bubble wall.  
However, for the symmetry-restoring scenario, the pressure is negative for small values of $\gamma$, which tends to accelerate the bubble wall.  
The possibility of this negative pressure had been noted in earlier work~\cite{Azatov:2024auq}, and we have demonstrated it with an explicit calculation. 
One can understand the negative pressure from the wall's longitudinal momentum change, which is also negative in a region of phase space.  
Note that $\Delta p_{1\to2} \approx (-\mah^2 + k_\perp^2/x) / 2E_a$ in \eref{eq:Dp_1to2_cases}, and since the integral over $k_\perp$ is dominated by the mass $k_\perp \approx \mbh$, and the integral over $x$ is dominated by the IR cutoff $x \approx x_\mathrm{min} = \mbh/E_a$, then it follows that $\Delta p_{1\to2} \sim (-\mah^2 + \mbh \gamma T) / 2E_a$. 
For small values of $\gamma$, the negative mass term dominates, which is why the 1-to-2 thermal pressure is negative for small $\gamma$.  
For large values of $\gamma$, the positive transverse momentum term dominates, and the pressure is positive. 
For intermediate values of $\gamma$, the pressure vanishes.  

\begin{figure}[t]
    \centering
    \includegraphics[width=0.70\linewidth]{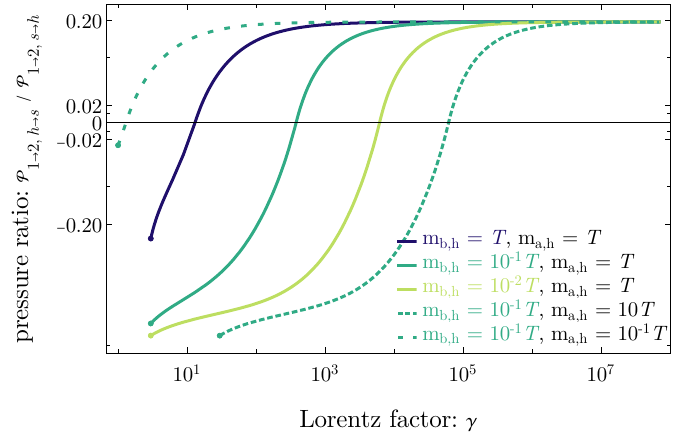}
    \caption{\label{fig:plot_ratio_sum}
    The ratio of the pressure in the symmetry-restoring scenario to that in the symmetry-breaking scenario as a function of the wall's Lorentz factor for several choices of masses. }
\end{figure}

\Fref{fig:plot_ratio_sum} shows the ratio of the 1-to-2 thermal pressure in the symmetry-restoring scenario to that in the symmetry-breaking scenario,  $\mathcal{P}_{1 \to 2, \mathrm{h} \to \mathrm{s}} /  \mathcal{P}_{1 \to 2, \mathrm{s} \to \mathrm{h}}$.  
For large values of the Lorentz factor $\gamma$, the ratio asymptotes to $\approx 0.2$ for all values of the masses that are shown.  
This universal scaling can be understood from the expression in \eref{eq:Dp_1to2_cases} for the wall's longitudinal momentum change $\Delta p_{1\to2}$.  
For small $x$ the leading terms are $\Delta p_{1\to2} \approx (\mbh^2 + k_\perp^2) / 2 x E_a$ for the symmetry-breaking scenario and $\Delta p_{1\to2} \approx k_\perp^2 / 2 x E_a$ for the symmetry-restoring scenario. 
The absence of the $\mbh^2$ term causes the symmetry-restoring scenario to have a pressure that is smaller by an order one factor since the integral is dominated by $k_\perp \sim \mbh$.  

We find that the curves in \Fref{fig:plot_ratio_sum} admit a simple empirical fitting function 
\begin{equation}
\begin{split}
    \frac{\mathcal{P}_{1 \to 2, \mathrm{h} \to \mathrm{s}}}{\mathcal{P}_{1 \to 2, \mathrm{s} \to \mathrm{h}}} 
    & \approx 0.5 \tanh\, [\, \log (\gamma/\gamma_\ast) + \mathrm{arctanh}(0.6)\,] - 0.3\,, \\ 
    \gamma_* 
    & \approx \frac{26 \, (\mah/T)^{2.2}}{(1 + |\log \mbh/T|^{1.65}) \, (\mbh/T)^{1.5}} - 12 \log(\mah/T) - 13
\end{split}
\label{eq:fit}
\end{equation}
for $\mbh < T, \mah$. 
This formula can be used to approximate the transition pressure in the symmetry-restoring scenario as a function of the masses $\mah$ and $\mbh$ in a specific model without performing any numerical integrals.

\subsection{Special Lorentz factors}
\label{sub:Lorentz}

Focusing now on the symmetry-restoring scenario, it is useful to identify two special values of the Lorentz factor $\gamma$.  
First we define $\gamma_\ast$ to be the value of $\gamma$ at which the pressure vanishes: 
\begin{align}
    \mathcal{P}_{1\to 2, \mathrm{h} \to \mathrm{s}}|_{\gamma_\ast} = 0 
    \;.
    \label{eq:gamma0}
\end{align}
Second we define $\gamma_{90}$ to be the value of $\gamma$ at which the the pressure ratio, which appears in \fref{fig:plot_ratio_sum}, has reached 90\% of its asymptotic value: 
\begin{align}
    & 
    \frac{\mathcal{P}_{1\to 2, \mathrm{h}\to \mathrm{s}}}{\mathcal{P}_{1\to 2, \mathrm{s}\to \mathrm{h}}} \biggr|_{\gamma_{90}} = 0.9 \times \lim_{\gamma\to\infty} \frac{\mathcal{P}_{1\to 2, \mathrm{h}\to \mathrm{s}}}{\mathcal{P}_{1\to 2, \mathrm{s}\to \mathrm{h}}}
    \;.
\label{eq:gamma90}
\end{align} 
Note that $0 < \gamma_\ast < \gamma_{90}$.  
The choice of 90\% is arbitrary. 
These two quantities, $\gamma_\ast$ and $\gamma_{90}$, represent two important stages in the bubble expansion process where the pressure changes qualitatively. 
Below $\gamma_\ast$, the pressure is negative, hence thermal pressure does not slow the wall down. 
Above $\gamma_\ast$, the pressure switches to positive values, hence the pressure acts as a frictional force, as familiar in the standard (symmetry-breaking) phase transitions. 
However, the pressure is smaller in magnitude than in the symmetry-breaking scenario, hence the frictional force is not as efficient at slowing down the bubble walls. 
The behavior familiar from the symmetry-breaking scenario, where pressure scales as $\sim \gamma \mbh T^3$, is achieved only at $\sim \gamma_{90}$. 
Beyond $\gamma_{90}$, we can expect the pressure to scale in the same way as the symmetry-breaking scenario for the same mass parameters. 

\begin{figure}[t]
    \centering
    \includegraphics[width=0.48\linewidth]{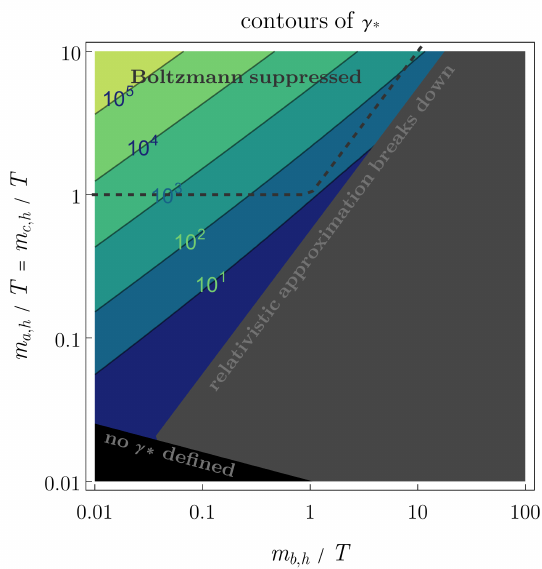} \hfill
    \includegraphics[width=0.48\linewidth]{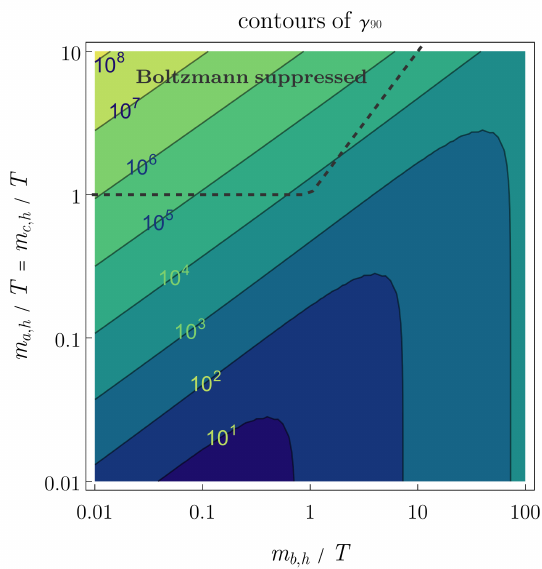}
    \caption{\label{fig:gammacontours}
    Special Lorentz factors across the parameter space of particle masses.  In both panels we show the parameter space in which the broken-phase masses ($\mah=\mch$ and $\mbh$) are varied.  The first panel shows contours of $\gamma_\ast$ (defined in \eref{eq:gamma0}), and the second panel shows $\gamma_{90}$ (defined in \eref{eq:gamma90}).  
    }
\end{figure}
In \fref{fig:gammacontours} we present two contour plots that show how $\gamma_\ast$ and $\gamma_{90}$ vary with the masses.  
For the plot of $\gamma_\ast$, in the bottom-left black region the thermal pressure does not vanish for any $\gamma$.   
For both plots, the top-left region bounded by the black-dashed lines, defined by the union of the two bounds $\mah > T$ and $\mah > \mbh$, represents the parameter space where the calculation is either inapplicable or requires additional model building.  
If $\mah \gtrsim T$ \textit{and} $\mah > \mbh$, then Boltzmann suppression (depletion of the $a$-particle abundance via $aa \to bb$ annihilations) would prevent $a$-particles from achieving a thermal abundance $n_a \sim T^3$, which was assumed in the derivation of the pressure.  
This constraint can be circumvented with additional model-building: for example, if $a$-particles dropped out of thermal equilibrium while they were still relativistic, then $aa\to bb$ annihilations are inactive at $T<\mah$, and a thermal abundance can be maintained even for $\mah>T$. 
Note that there is no analogous restriction on $\mbh > T$, since the pressure calculation only assumes a thermal population of $a$-particles, and the $b$-particles are not required to be present in the thermal bath.  
Recall that various approximations used in the B\"odeker and Moore calculation break down if $\gamma_\ast T \lesssim \mah, \mbh$; in the top panel, this region is represented by the bottom grey shaded region, where we have used $\gamma_\ast T > 3 \mbh$ as a conservative boundary. 

\subsection{Terminal velocity}
\label{sub:terminal_velocity}

Relativistic bubble walls can reach a terminal velocity at which the thermal pressure is balanced against the vacuum pressure~\cite{Bodeker:2017cim}.  
For this discussion, we denote the thermal pressure by $\mathcal{P}_\mathrm{th} \approx \mathcal{P}_{1\to1} + \mathcal{P}_{1\to2}$, and we denote the vacuum pressure by $\mathcal{P}_\mathrm{vac} = - \Delta V$ where $\Delta V$ is the latent heat of the phase transition. 
If $P_\mathrm{vac} + P_\mathrm{th} < 0$ then the bubble wall accelerates and the wall's Lorentz factor $\gamma$ grows larger.  
We have discussed that $\mathcal{P}_{1\to1} \propto \gamma^0$ and that $\mathcal{P}_{1\to2} \propto \gamma^1$ for large $\gamma$.  
We also note that $\mathcal{P}_\mathrm{vac} \propto \gamma^0$.  
So as the wall accelerates, the bubble wall eventually reaches a terminal velocity at which $\gamma = \gamma_t$ and $P_\mathrm{vac} + P_\mathrm{th} = 0$.   

\begin{figure}[t]
    \centering
    \includegraphics[width=0.70\linewidth]{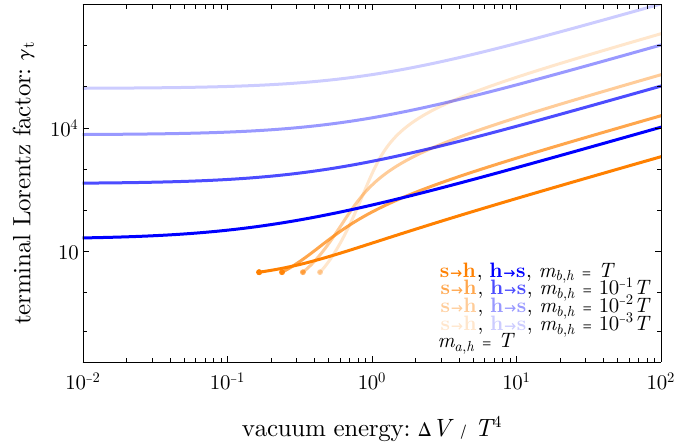} 
    \caption{\label{fig:gamma_t}
    Terminal Lorentz factor $\gamma_t$ for various choices of particle masses for symmetry-restoring scenarios ($\mathrm{h}\to \mathrm{s}$, blue curves) and symmetry-breaking scenarios ($\mathrm{s}\to \mathrm{h}$, orange curves) as a function of the difference in vacuum energy $\Delta V / T^4$. We take $g^2 C_2[R] = 1$. 
    }
\end{figure}

In \Fref{fig:gamma_t} we show how the terminal Lorentz factor $\gamma_t$ depends on the vacuum energy $\Delta V$ for several choices of the particle masses for symmetry-breaking (orange) and symmetry-restoring (blue) transitions.    
For large values of $\Delta V$, we observe that the terminal Lorentz factor grows as $\gamma_t \propto \Delta V$ for both the symmetry-breaking and symmetry-restoring scenarios.  
However, the bubble walls move more quickly in the symmetry-restoring scenario, and their $\gamma_t$ is larger by a factor of $\approx 10$.  

We see that this implies that the terminal behavior is reached for $\gamma\sim  \mathcal{O}(10-10^5)$ for the shown masses in the symmetry-breaking scenario. 
For smaller values of $\Delta V$, we observe that $\gamma_t$ decreases sharply for $\Delta V / T^4 \lesssim 1$ in the symmetry-breaking scenario.  
From comparing with \Fref{fig:plot_all_contributions} we see that in the symmetry-breaking scenario the pressure reaches a plateau for low $\gamma$, which comes from the positive contributions from $\mbh$ and $\mch$ appearing in the momentum transfer in \eref{eq:P_1to2_simple}. 
This plateau means that a small increase in $\Delta V$ leads to a large increase in $\gamma_t$. 
However, in the symmetry-restoring scenario the terminal Lorentz factor remains large even as $\Delta V / T^4 \ll 1$.  
This is because of the negative contributions for small $\gamma$ in the symmetry-restoring scenario. 
Since the thermal pressure drops to zero at finite $\gamma = \gamma_\ast$, an arbitrarily small $\Delta V$ will not correspond to an arbitrarily small $\gamma_t$, but rather a finite $\gamma_t \sim \gamma_\ast$.  

\subsection{Implications for phenomenology}
\label{sub:pheno}

Various cosmological relics are expected to arise from first-order cosmological phase transitions.  
These include gravitational wave radiation, primordial magnetic fields, particle asymmetries such as the cosmological baryon asymmetry, stable particles such as dark matter, topological defects, and primordial black holes.  
The detection of these relics in the universe today is of paramount importance for understanding the physics of the early universe.  
We have seen that first-order phase transitions in the symmetry-restoring scenario tend to produce faster bubbles than the symmetry-breaking scenario.  
Here we briefly remark on a few consequences and observable signatures of these faster bubbles.  

\vskip 0.1cm 
\noindent\textit{Gravitational Waves:} 

A stochastic background of gravitational wave  radiation is expected to arise from a cosmological first-order phase transition. If the bubble walls achieve runaway behavior, GWs are sourced by the scalar field energy densities during bubble collision~\cite{Kosowsky:1991ua,Kosowsky:1992rz,Kosowsky:1992vn,Kamionkowski:1993fg,Caprini:2007xq,Huber:2008hg,Bodeker:2009qy,Jinno:2016vai,Jinno:2017fby,Konstandin:2017sat,Cutting:2018tjt,Cutting:2020nla,Lewicki:2020azd} or by the distribution of particles produced from bubble collisions~\cite{Inomata:2024rkt}. On the other hand, if the bubble walls achieve some terminal velocity long before the bubbles collide, GWs instead arise from sound waves~\cite{Hindmarsh:2013xza,Hindmarsh:2015qta,Hindmarsh:2017gnf,Cutting:2019zws,Hindmarsh:2016lnk,Hindmarsh:2019phv} and turbulence~\cite{Kamionkowski:1993fg,Caprini:2009yp,Brandenburg:2017neh,Cutting:2019zws,RoperPol:2019wvy,Dahl:2021wyk,Auclair:2022jod} in the plasma, or through nontrivial spatial configurations of particles that do not thermalize~\cite{Jinno:2022fom}. The GWs produced by these various sources have different peak frequencies and spectral shapes (e.g.\,see Fig.\,3 in \Rref{Inomata:2024rkt}), as determined by the details of the underlying physics. 

Since a symmetry-restoring phase transition tends to reach larger Lorentz factors, a greater fraction of the energy released in the phase transition is stored in the bubble walls at collision. This implies that the GW contribution from the first group of sources above (bubble collision, particles produced from collisions) is always greater in a symmetry-restoring transition. The amplitude of this GW component is expected to scale as the square of the energy density on the bubble walls, which scales linearly with $\gamma_t$; hence $\Omega_{GW}\propto \gamma_t^2$ for this component. The GW signal from symmetry-breaking and restoring transitions are therefore expected to be different even in the same theory (i.e.\,with the same particles and similar $\Delta V$) (see also \Rref{Buen-Abad:2023hex}).

\vskip 0.1cm
\noindent\textit{Heavy Relics:} 

During a first-order cosmological phase transition, heavy particles can be produced even if their mass is far above the temperature of the phase transition~\cite{Azatov:2021ifm,Baldes:2021vyz,Azatov:2021irb,Baldes:2022oev,Chun:2023ezg,Mansour:2023fwj,Shakya:2023kjf,Ai:2024ikj,Giudice:2024tcp,Cataldi:2024pgt,Cataldi:2025nac}. This is possible because the energy scale of bubble collisions or bubble-plasma interactions is set by the boosted bubble wall energy $\sim \gamma T$, which can be several orders of magnitude higher than the temperature when $\gamma_t\gg1$. Such interactions can produce massive states with mass $T<m<\gamma_t T$. 

Since the symmetry-restoring transitions always reach larger $\gamma_t$ values compared to the symmetry-breaking counterpart, they are more amenable to the production of such heavy states. For concreteness, consider the case where $\mah = T, \mbh = 10^{-3} T$ and $\Delta V/T^4 = 1$ in \Fref{fig:gamma_t}. Then $\gamma_t\sim 10^3 (10^5)$ for the symmetry-breaking (restoring) scenario. In this case, states with mass $m\approx 10^4 T$ can be readily produced in the symmetry-restoring scenario, but not the symmetry-breaking scenario. Such particles could be dark matter~\cite{Azatov:2021ifm,Baldes:2022oev,Ai:2024ikj,Giudice:2024tcp}, source the baryon asymmetry of the Universe~\cite{Baldes:2021vyz,Azatov:2021irb,Chun:2023ezg,Cataldi:2024pgt,Cataldi:2025nac}, or be of interest for other reasons. In this sense, the phenomenology of symmetry-restoring transitions can be far richer than that of the symmetry-breaking ones due to the larger terminal Lorentz factors reached by the bubble walls. 

\section{Summary}
\label{sec:summary}

We are interested in the dynamics of ultrarelativistic bubble walls during a cosmological first-order phase transition in the symmetry-restoring scenario.  
We calculate the thermal pressure that arises when particles in the plasma interact with the bubble wall. 
Earlier work by other authors has established that the 1-to-1 contribution to the thermal pressure is negative, and that the 1-to-2 contribution is positive in the limit where the wall's Lorentz factor $\gamma$ is large.  
In our work, we are particularly interested in understanding how the 1-to-2 thermal pressure behaves for small and intermediate values of $\gamma$.  In the remainder of this summary, we highlight several qualitative results that are important for various phenomenological applications.   
\begin{itemize}

\item The main result of our study is that the 1-to-2 thermal pressure can be \textit{negative} if the bubble wall's Lorentz factor $\gamma < \gamma_\ast$ takes small-to-intermediate values (see \Fref{fig:plot_all_contributions}). 
This is in contrast with symmetry-breaking transitions, where the 1-to-2 pressure is always positive. The left panel of \Fref{fig:gammacontours} shows how $\gamma_\ast$ depends on the masses of particles in our model (in which the 1-to-2 pressure is dominated by the emission of a transversely-polarized vector boson). This result is contrary to the intuition that the presence of significant interactions between the expanding bubble walls and the surrounding plasma produces a frictional force that slows the walls down. 

\item For large values of $\gamma \gtrsim \gamma_{90}$, we find that the 1-to-2 thermal pressure is positive and scales $\propto \gamma$, consistent with the results reported in \Rref{Azatov:2024auq}.  
The right panel of \Fref{fig:gammacontours} shows how $\gamma_{90}$ depends on the particle masses. 

\item We provide a simple empirical fit function \pref{eq:fit} for the 1-to-2 thermal pressure as a function of the masses involved. 
This can be used to obtain an estimate of the pressure in a specific model of symmetry-restoring transition in a straightforward manner without performing 
numerical phase-space integrations. 

\item  We calculate the Lorentz factor $\gamma_t$ corresponding to the terminal velocity at which the thermal pressure balances the vacuum pressure.  Since the thermal pressure in the symmetry-restoring scenario is smaller than the pressure in the symmetry-breaking scenario, it follows that the terminal velocity is larger.  This enhancement factor is found to be $\gtrsim 10$ for the parameters illustrated in \Fref{fig:gamma_t}.  

\item The relatively larger terminal boost factor in a symmetry-restoring transition can have important phenomenological implications and applications. It can enhance the component of the gravitational wave signal produced from bubble collisions and from particles produced at bubble collisions, qualitatively modifying the peak, amplitude, and spectral features of the overall gravitational wave signal. Similarly, it can also enhance the production of particles with masses far higher than the temperature of the plasma, opening rich possibilities for various beyond-the-Standard-Model applications such as the production of dark matter and the baryon asymmetry of the Universe.    

\end{itemize}

\section*{Acknowledgements}
We are grateful to Aleksandr Azatov, Simone Blasi, Thomas Konstandin, Gudrid Moortgat-Pick, Yikun Wang, and Miguel Vanvlasselaer for helpful discussions. This work was supported by the Deutsche Forschungsgemeinschaft under Germany's Excellence Strategy - EXC 2121 ``Quantum Universe" - 390833306. A.J.L. is partly supported by the NASA ATP award 80NSSC22K0825. This research was supported by the Munich Institute for Astro-, Particle and BioPhysics (MIAPbP) which is funded by the Deutsche Forschungsgemeinschaft (DFG, German Research Foundation) under Germany´s Excellence Strategy – EXC-2094 – 390783311. 

\bibliographystyle{JHEP}
{\footnotesize
\bibliography{FOPTreferences}

\end{document}